\documentstyle[12pt]{article}
\newcommand{\bm}[1]{\mbox{\boldmath $#1$}}

\title{ON THE CAUSAL PROPAGATION OF FIELDS}  
\author{G\"{o}ran Bergqvist \\ Department of Mathematics,
 M{\"a}lardalen University \\ 721 23 V{\"a}ster{\aa}s, Sweden
(gbt@mdh.se) \\ \\ Jos\'{e} M. M. Senovilla \\
Departamento de F\'{\i}sica Te\'orica, Universidad del Pa\'{\i}s Vasco\\
Apartado 644, 48080 Bilbao, Spain \\  and \\
Departament de F\'{\i}sica Fonamental,
 Universitat de Barcelona \\ Diagonal 647, E-08028 Barcelona, Spain \\
(seno@ffn.ub.es)}
\begin{document}           

\maketitle                 

\abstract
By using geometric methods and superenergy tensors,
we find new simple criteria for the causal propagation of
physical fields in spacetimes of any dimension. The method can be applied
easily to many different theories and to arbitrary fields (such as scalar
or electromagnetic ones). In particular, it provides a conservation theorem
of the free gravitational field in all $N$-dimensional spacetimes conformally
related to Einstein spaces (including the Ricci-flat case). In 
general relativity, our criteria provide simple proofs and a unified
treatment of conservation theorems for neutrinos, photons, electrons and
all other massless and massive free spin $n/2$ fields. The uniqueness of
the solution to the field equations also follows from our treatment
under certain circumstances.

\vspace{1cm}

\newpage

The traditional way of studying general properties of
field equations for physical fields on spacetimes is to
introduce some coordinates and translate
the geometric structure into the language of standard
theory of partial differential equations to apply analytic
methods or to consider the existence of
fundamental solutions (propagators), see
\cite{Courant-Hilbert,Friedlander,Ch,F,ACY,Reula,Rendall,Bo}
and references therein.
In this letter we put forward a way to use the geometric information
of tensorial field equations
in a more direct manner. We show how a conservation theorem
for fields can be proven, a theorem which provides information
about the causal propagation of the fields.
Our methods work for spacetimes of any dimension, so that it can be applied to
many different theories including string theory, M-theory (see e.g. \cite{Gib}),
supergravity, gauge theories, Kaluza-Klein-like theories, Brans-Dicke and
scalar-tensor theories, dilaton gravity and many others.
The idea is to generalize the conservation theorem proved in \cite{HE} for
the energy-momentum tensors that satisfy the dominant energy condition to
the case of arbitrary physical fields. The key points in our reasoning are:
first, ``superenergy" tensors with the corresponding dominant
property can {\it always} be constructed and, second, these tensors
do not have to be related to any physically meaningful quantity, nor to the
energy-momentum tensor of the field (which, in fact, cannot be used
in many cases, see below, including the important case of the gravitational
field where it cannot be defined).
The {\it only} condition in our conservation theorem is an assumption on the
divergence of the superenergy tensor which is easy to check
explicitly.

Our method is clearly related to the techniques of symmetric
hyperbolic systems. Friedrich \cite{F,F2} has used these techniques to
study many interesting equations including several of the examples
we will discuss here. The great advantage of our method is that we
do not need to put the equations on a form which fits the theory
of symmetric hyperbolic systems but we use the superenergy tensor directly
in a typical integration procedure and to obtain estimates.

Let spacetime $V_N$ be an $N$-dimensional Lorentzian manifold
with a metric $g$ of signature (+, --, ..., --). A rank-$m$ tensor $T$
is said to satisfy the {\it dominant property} if \cite{S1,B2,S3}
\begin{eqnarray*}
T_{a_1\dots a_m}u_1^{a_1}\dots u_m^{a_m}\ge 0
\end{eqnarray*}
for any set $\{ \vec{u}_1,\dots ,\vec{u}_m\}$ of causal future-pointing
vectors. This is equivalent \cite{S3,S1} to that with respect to an orthonormal
basis $\{\vec{e}_0,\dots ,\vec{e}_{N-1}\}$ the pure time component dominates
any other component, that is
\begin{equation}
T_{0\dots 0}\ge |T_{\alpha\dots\beta}| \label{dp2}
\end{equation}
for any $0\le\alpha,\dots ,\beta\le N-1$. Note that this also
implies that $T_{0\dots 0}=0 \Rightarrow  T=0$.
(Observe that the dominant property is a generalization of the dominant
energy condition for rank-2 energy-momentum tensors \cite{HE}.)

In general, a physical field $A$ will not have the dominant property
but, as has been proven in complete generality \cite{S1,B2} and in arbitrary
dimensions \cite{S3}, one can {\it always}
construct superenergy tensors $T(A)$ of $A$ which have the dominant
property and such that, in any open neighbourhood,
\begin{equation}
A=0 \Longleftrightarrow T(A)=0 \, .
\label{eq:A0T0}
\end{equation}
This property holds in open neighbourhoods of the manifold, but it may need a
refinement at isolated points or lower-dimensional subsets. For instance,
sometimes $T(A)$ vanishes at a point if and only if $A$ {\it and} its
derivatives up to a certain order are zero \cite{S3}, see for an explicit clear
example the case of the massive scalar field below. We shall denote by ${\cal A}$
the set formed by the tensor $A$ and its derivatives up to the required order
such that, at any point $x\in V_N$
\begin{equation}
{\cal A} |_x=0 \Longleftrightarrow T(A) |_x=0 \, .
\label{eq:A0T0'}
\end{equation}
The set ${\cal A}$ essentially contains the derivatives of $A$ appearing
explicitly in the expression of $T(A)$ for each case.

The construction of $T(A)$
generalizes the definitions of energy-momentum tensors of scalar and Maxwell
fields \cite{S1,B2,S3} and the traditional Bel and Bel-Robinson superenergy
tensors of the gravitational field \cite{S1,B2,S3,PR,BoS1,B1}. Many examples of
$T(A)$ will be given below. Note that the standard energy-momentum tensors of
physical fields, even in the cases they are well defined, do not always
have the dominant property \cite{PR} so they cannot be used as the tensor
$T(A)$ in general for our purposes.

Let us try to see now under what conditions the physical field $A$ will
propagate causally. To that end, assume that $S$ is an
$(N-1)$-dimensional closed achronal set in $V_N$ and $D^+(S)$ its future
Cauchy development (see \cite{HE} for notation and definitions).
As $S$ is achronal, int$D^+(S)$ is globally hyperbolic and it can therefore be
foliated by spacelike hypersurfaces $\Sigma_t\equiv \{ t=$const.$\}$,
where the gradient $\bm{v}=d t$ is timelike everywhere in int$D^+(S)$.
Take any point $q\in D^+(S)$, then $K\equiv J^-(q)\cap\overline{D^+(S)}$
is compact and has boundary $\partial K=\tilde S \cup H^+(\tilde S)$,
where $\tilde S\equiv S\cap J^-(q)$ and $H^+(\tilde S)$ is the
future Cauchy horizon of $\tilde S$.
Using any of the superenergy tensors $T(A)$ we define the superenergy integral
\begin{eqnarray}
w(t)\equiv \int_{J^-(\Sigma_t)\cap K}T^{a_1\dots a_m}(A)
v_{a_1}\dots v_{a_m} \bm{\eta}
=\int_{t_0}^t \left( \int_{\Sigma_{t'}}T^{a_1\dots a_m}(A)
v_{a_2}\dots v_{a_m}d\sigma_{a_1}|_{\Sigma_{t'}} \right) dt'
\label{eq:w}
\end{eqnarray}
Here $J^-(\Sigma_t)$ denotes the causal past of $\Sigma_t$,
$\bm{\eta}$ the canonical volume $N$-form on $V_N$,
$d\sigma_{a_1}|_{\Sigma_{t'}}$ the volume $(N-1)$-form on
$\Sigma_{t'}$ pointing along $\vec{v}$, and $t_0$ the minimal value
of $t$ in $K$.

The procedure now follows
the structure of a theorem for energy-momentum tensors
in \cite{HE}.
Differentiating, we get by Gauss' theorem
\begin{eqnarray}
\frac{dw}{dt}\equiv w'(t)= \int_{\Sigma_{t}}T^{a_1\dots a_m}(A)
v_{a_2}\dots v_{a_m}d\sigma_{a_1}|_{\Sigma_{t}}= \hspace{2cm} \label{eq:wp} \\
=-\int_{H^+(\tilde S)\cap J^-(\Sigma_{t})}T^{a_1\dots a_m}(A)
v_{a_2}\dots v_{a_m}d\sigma_{a_1}|_{H^+(\tilde S)}
-\int_{\tilde S}T^{a_1\dots a_m}(A)
v_{a_2}\dots v_{a_m}d\sigma_{a_1}|_{\tilde S}+ \nonumber  \\
+\int_{J^-(\Sigma_t)\cap K}
\left\{[\nabla_{a_1}T^{a_1\dots a_m}(A)]v_{a_2}\dots v_{a_m} +\right.
\hspace{2cm}\label{eq:wp2} \\
\left.+T^{a_1\dots a_m}(A)\left[(\nabla_{a_1}v_{a_2})v_{a_3}\dots v_{a_m}
+\dots + v_{a_2}\dots v_{a_{m-1}}(\nabla_{a_1}v_{a_m})\right]\right\}
\bm{\eta}
\nonumber
\end{eqnarray}
where $d\sigma_{a}|_{\zeta}$ always denotes the outward-pointing volume
$(N-1)$-form on any $(N-1)$-dimensional subset $\zeta$. Given that $T(A)$ has
the dominant property, by (\ref{eq:w}) and (\ref{eq:wp}) we have 
\begin{eqnarray*}
w(t)\ge 0, \hspace{1cm} w'(t)\geq 0 \, .
\end{eqnarray*}
By (\ref{dp2}) we get that
$T^{a_1\dots a_m}(A)(\nabla_{a_1}v_{a_2})v_{a_3}\dots v_{a_m}
\le M_2 T^{a_1\dots a_m}(A) v_{a_1}\dots v_{a_m}$ on
$K$ for some constant $M_2$, because $\nabla_{a_1}v_{a_2}$
is bounded on the {\it compact} $K$.
Similarily, we get constants $M_3, \dots ,M_m$ from the other terms so that
putting $M\equiv M_2+\dots +M_m$ and using the fact that the normal to
$H^+(\tilde S)$ pointing outwards from $K$ is future-pointing, we get by
(\ref{eq:wp2}) that
\begin{eqnarray*}
0\!\le \! w'(t)\! \le \!Mw(t)
\!-\!\!\int_{\tilde S}T^{a_1\dots a_m}(A)
v_{a_2}\dots v_{a_m}d\sigma_{a_1}|_{\tilde S}
\!+\!\!\int_{J^-(\Sigma_t)\cap K}
[\nabla_{a_1}T^{a_1\dots a_m}(A)]v_{a_2}\dots v_{a_m}  \bm{\eta} \, .
\end{eqnarray*}
If ${\cal A}=0$ on $\tilde S$ (we usually think of $S$ as an initial
slice), then due to the property (\ref{eq:A0T0'}) of the superenergy tensors
$T(A)$ we obtain
\begin{equation}
0\le w'(t)\le Mw(t)+
\int_{J^-(\Sigma_t)\cap K}
[\nabla_{a_1}T^{a_1\dots a_m}(A)] v_{a_2}\dots v_{a_m}  \bm{\eta}\, .
\label{eq:wp2b}
\end{equation}
Consider now the following {\it divergence condition} \footnote{This is
enough for our purposes, but it can be obviously generalized to other more
general conditions involving ``powers'' of the tensor $T$ greater than one.
On the other hand, (\ref{eq:div}) can be written in a form explicitly
independent of the foliation, but this is not needed here.}:
\begin{equation}
\nabla_{a_1}T^{a_1\dots a_m}(A) v_{a_2}\dots v_{a_m}\le
h\,T^{a_1\dots a_m}(A) v_{a_1}\dots v_{a_m}
\label{eq:div}
\end{equation}
where $h$ is a continuous function on $\overline{D^+(S)}$. Then
eq.(\ref{eq:wp2b}) becomes simply
\begin{eqnarray*}
0\le w'(t)\le (M+H) w(t)
\end{eqnarray*}
where the constant $H$ is the maximum of $h$ on
$K$. Integrating this relation (or using a rather simple Gronwall's inequality)
and using $w(t_0)=0$, one then arrives at
\begin{equation}
w(t)\equiv 0 \, .
\label{eq:w0}
\end{equation}
This means that $T_{0\dots 0}(A)\equiv 0$ on $K$ (with
$\vec{e}_0\propto \vec{v}$)
and hence $T(A)\equiv 0$ on $K$. For any point $q\in D^+(S)$
we can apply the above reasoning to the corresponding compact
$K$ to obtain $T(A)=0$ on $K$ and hence we get that $T(A)=0$
on $\overline{D^+(S)}$ by continuity,
which is equivalent to $A=0$ on $K$ due to (\ref{eq:A0T0}). Note that changing
future to past we also get $A=0$ in $\overline{D^-(S)}$ and hence
in $\overline{D(S)}=\overline{D^+(S)\cup D^-(S)}$.

Using the terminology of \cite{HE}, we have thus proven the following
\smallskip

{\it Conservation theorem:
${\cal A}|_S=0 \Longrightarrow A|_{\overline{D(S)}}=0$
if the superenergy tensor $T(A)$ satisfies (\ref{eq:div}).}

\smallskip

This theorem is clearly related to the causal propagation of the field
described by $A$, because if $A\neq 0$ in any region arbitrarily close
to $\overline{D(S)}$ then $A$ will propagate in time according to the
field equations but it will never be able to ``enter'' $\overline{D(S)}$,
showing that $A$ cannot travel faster than light. On the other hand, the
universality of the superenergy construction \cite{S1,B2,S3} is essential
here.

As an example in arbitrary dimension, consider a 
scalar field $\phi$ with mass $m$ satisfying the `generalized Klein-Gordon'
$N$-dimensional equation $\left(\Box +m^2\right) \phi=k^a\nabla_a\phi$,
where $k^a$ is some continuous vector field.
Define the tensor $T(\phi )$ as
\begin{eqnarray*}
T_{ab}=\nabla_a\phi \nabla_b\phi-{1\over 2}g_{ab}\nabla_c\phi\nabla^c\phi
+\frac{1}{2}m^2 \phi^2 g_{ab}
\end{eqnarray*}
(this is the standard energy-momentum tensor {\it if} $k^a=0$, in which case
$T(\phi )$ is divergence-free). $T(\phi )$ is easily shown to have the dominant
property in any dimension and to be zero if and only if $\phi =0$
(iff $\nabla_a\phi =0$ for the massless case) in open neighbourhoods \cite{S3}.
However, at points $x\in V_N$, we have \cite{S3}
\begin{eqnarray*}
T(\phi )|_x =0 &\Longleftrightarrow & \phi |_x =0, \, \, \nabla\phi |_x =0
\hspace{1cm} \mbox{if} \,\, m\neq 0 \\
T(\phi )|_x =0 &\Longleftrightarrow & \nabla\phi |_x =0
\hspace{2cm} \mbox{if} \,\, m= 0
\end{eqnarray*}
Thus, in this case the set ${\cal A}$ is given by ${\cal A}\equiv
\left\{\phi,\nabla\phi\right\}$ in the massive case and by ${\cal A}\equiv
\nabla\phi$ in the massless one. In any case,
$T(\phi )$ satisfies the divergence condition (\ref{eq:div}) so that the
theorem shows that $\phi|_S , \nabla\phi |_S =0 \Longrightarrow
\phi |_{\overline{D(S)}}=0$ if $m\neq 0$,
($\nabla\phi|_S =0\Longrightarrow \phi |_{\overline{D(S)}}=$const. if $m=0$).
In addition, due to the linearity of the generalized Klein-Gordon equation, this
result implies that $\phi |_{\overline{D(S)}}$ is {\it unique} given
the initial condition $\phi|_S,\nabla\phi |_S$. 
A similar proof works for the electromagnetic field using the energy-momentum
tensor in dimension $N$.

Much more interesting is the case of the free gravitational field in $N$
dimensions, because then {\it no} energy-momentum tensor exists (due to the
Equivalence Principle and the non-localizability of the gravitational energy).
However, one can use the superenergy tensors as follows.
Let ${\cal T}$ be the generalized Bel-Robinson tensor given by \cite{S3}
\begin{eqnarray*}
{\cal T}_{abcd}\!=\!C_a{}^e{}_c{}^f C_{bedf}\!+\!C_a{}^e{}_d{}^f C_{becf}
\!-\!\frac{1}{2}g_{ab}C^{le}{}_c{}^f C_{ledf}
\!-\!\frac{1}{2}g_{cd}C_a{}^e{}^{lf} C_{belf}\!+\!
\frac{1}{8}g_{ab}g_{cd}C^{helf} C_{helf},\\
{\cal T}_{abcd}={\cal T}_{bacd}={\cal T}_{abdc}={\cal T}_{cdab}
\hspace{5cm} \\
{\cal T}^a{}_{acd}=(N-4){\cal T}^a{}_{cad}, \hspace{5mm}
{\cal T}^a{}_{cad}=-\frac{1}{2}\left(C_{clef}C_d{}^{lef}
-\frac{1}{4}g_{cd}C^{helf} C_{helf}\right)
\end{eqnarray*}
where $C$ is the Weyl tensor. Then
${\cal T}$ has the dominant property \cite{S3} (see \cite{PR,BoS1,B1}
for $N$=4, in which case ${\cal T}$ is completely symmetric and traceless)
and vanishes at any point if and only if $C$ is zero there \cite{S3}.
Let $\tilde g$ be any metric conformally related to $g$ by
\begin{equation}
\tilde g =e^{2U}g
\label{eq:divC}
\end{equation}
Then $\tilde C^a{}_{bcd}=C^a{}_{bcd}$ and
\begin{equation}
\tilde\nabla_a \tilde C^a{}_{bcd}=\nabla_a C^a{}_{bcd}+(N-3)
U_{,a} C^a{}_{bcd}
\label{eq:conf}
\end{equation}
which implies
\begin{eqnarray}
\tilde\nabla_a \tilde {\cal T}^a{}_{bcd}=
e^{-2U}\left[\nabla_a{\cal T}^a{}_{bcd}+
(N-4) U_{,a} {\cal T}^a{}_{bcd}+U_{,a}({\cal T}^a{}_{cbd}+
{\cal T}^a{}_{dbc})-\right.\nonumber\\
\left.-(N-4)U_{,b}{\cal T}^a{}_{cad}-U_{,c}{\cal T}^a{}_{bad}-
U_{,d}{\cal T}^a{}_{bac}\right].
\label{eq:divBR}
\end{eqnarray}
If $\nabla_a C^a{}_{bcd}=0$, then $\nabla_a {\cal T}^a{}_{bcd}=0$ and
by (\ref{eq:divBR}) we see that
the divergence condition holds for $\tilde {\cal T}$. This comprises
all spacetimes conformally related to Einstein spaces
($R_{ab}=\Lambda g_{ab}$), including the important case of Ricci-flat
metrics. Thus we have a simple proof of the causal
propagation of free gravity (the Weyl tensor) in all such spacetimes,
which are of big generality. The special case of General Relativity (where
$N=4$ and the Bel-Robinson tensor is divergence-free) was studied in \cite{BoS1}.

In the case of General Relativity, for many fields one can
construct a tensor $T$ satisfying (\ref{eq:div}). Note that
the energy-momentum tensors, which would be obvious candidates,
are not always well defined (as happens for massless fields of spin
higher than one), and that in many cases, such as for instance spin-$1/2$
fields, the standard energy-momentum tensors do not satisfy
the dominant energy condition \cite{PR}. Therefore, we must
construct some other tensor $T$ in these cases
in the same way as with the generalized Bel-Robinson tensor above.
As mentioned before, in fact one can
always construct a superenergy tensor $T(A)$ of any field $A$
with the dominant property \cite{S1,B2,S3}.

Consider a massless spin $n/2$ field $\varphi$. This is described by a
symmetric
spinor field $\varphi_{A_1 \dots A_n}=\varphi_{(A_1 \dots A_n)}$
(spinor indices are capital letters; see \cite{PR} for notation) and
the field equation is \cite{PR}
\begin{equation}
\nabla^{A_1 B'}\varphi_{A_1 \dots A_n}=0\, .
\label{eq:mlfe}
\end{equation}
For the neutrino ($n=1$) and the photon ($n=2$) this is a
well-posed initial value problem, but for $n>2$ there arise the Buchdahl
algebraic constraints \cite{Buc} involving the curvature for existence of
solutions. We refer to \cite{Buc,PR} for a discussion on these restrictions.
Nonetheless, independetly of these restrictions, we can apply our method to
prove causal propagation of the solutions in general.

Define the tensor $T(\varphi )$ by
\begin{eqnarray*}
T_{a_1 \dots a_n}=\varphi_{A_1 \dots A_n}\bar\varphi_{A'_1 \dots A'_n}
\end{eqnarray*}
(with the usual identification $a_1\leftrightarrow A_1A'_1$ between tensor and
spinor indices \cite{PR} and where the bar denotes complex conjugation).
As shown in \cite{B2}, any sum of spinors squared in this sense has the
dominant property. Note that for $n=2$ this is the
usual energy-momentum for Maxwell fields.
By (\ref{eq:mlfe}), $\nabla^{a_1}T_{a_1 \dots a_n}=0$
and as $T(\varphi )|_x=0 \Leftrightarrow \varphi |_x=0$,
the conservation theorem is valid for $\varphi$.
The theorem also holds for more general field equations.
In analogy with the generalized Klein-Gordon $N$-dimensional equation,
we can consider the simple generalization
\begin{eqnarray*}
\nabla^{A_1 B'}\varphi_{A_1 \dots A_n}=k^{A_1 B'}\varphi_{A_1 \dots A_n}
\end{eqnarray*}
for some continuous vector field $k^a$. Then the divergence condition
is also satisfied and therefore the theorem is applicable to
such fields too. In any case, the uniqueness of $\varphi$ on $\overline{D(S)}$
for a given initial condition $\varphi |_S$ holds.

For massive spin $n/2$ fields, one can find consistent systems of equations
(without any Buchdahl's constraints) for spinor fields $\varphi$, $\chi$
such as \cite{Bu},\cite{I}
\begin{equation}
\nabla^{A_1}{}_{B'}\varphi_{A_1 \dots A_n}=\mu\chi_{A_2 \dots A_n B'},
\hspace{1cm}
\nabla_{(A_1}{}^{B'}\chi_{A_2 \dots A_n)B'}=\nu\varphi_{A_1 \dots A_n}
\label{eq:Bu}
\end{equation}
with $\mu$, $\nu\ne 0$ and $m^2=2\mu\nu$. This is a well posed
initial value problem for all spins. Let
\begin{equation}
T_{a_1 \dots a_n}=\alpha\varphi_{A_1 \dots A_n}\bar\varphi_{A'_1 A'_2
 \dots A'_n}
+\beta\chi_{A'_1 A_2 \dots A_n}\bar\chi_{A_1 A'_2 \dots A'_n}
\label{eq:T}
\end{equation}
with constants $\alpha$, $\beta >0$.
Then $T$ has the dominant property, $T |_x=0 \Leftrightarrow
\varphi |_x=0=\chi |_x$ and
\begin{eqnarray*}
\nabla^{a_1}T_{a_1 \dots a_n}=
(\beta\nu-\alpha\mu)\chi_{A'_1 A_2 \dots A_n}\bar\varphi^{A'_1}{}_{A'_2
 \dots A'_n} + \hspace{3cm} \\
+{{\beta n(n-1)}\over{\mu (n+1)}}
\epsilon^{A_1}{}_{(A_2}\Psi_{A_3}{}^{DEF}
\varphi_{A_4\dots A_n)DEF}\bar\chi_{A_1 A'_2 \dots A'_n}
+\mbox{complex  conjugates.}
\end{eqnarray*}
Here we have used eq.(C1) of \cite{I} and $\Psi$ denotes the Weyl spinor.
For any $\alpha$ and $\beta$, any product of two components of
$\chi$ and $\varphi$ with a bounded coefficient will be smaller
than a constant times the time component $T_{0\dots 0}$ of $T$
(so it is not necessary to make the natural choice $\beta\nu=\alpha\mu$).
Thus the divergence condition (\ref{eq:div}) is satisfied and
$(\varphi , \chi )$ propagate causally. The uniqueness of $(\varphi , \chi )$
on $\overline{D(S)}$ given the initial condition $(\varphi , \chi )|_S$
follows again.

It is important to remark that the fact that $\mu$ is a constant is fundamental
in the above reasoning, because it allows to isolate the derivative of $\chi$
(eq.(C1) in \cite{I}) which in turn is used to check the divergence condition.
Thus, for instance, the case of a massless spin-$3/2$ field satisfies the
Rarita-Schwinger equations which, due to the Buchdahl constraints mentioned
before, must be modified in curved spacetime in an appropriate manner, see
e.g.\cite{PR,S-O}. The resulting equations in Ricci-flat spacetimes are formally
similar to (\ref{eq:Bu}) with $n=3$, {\it but} instead of the
constant $\mu$ there appears the Weyl spinor $\Psi$ contracted appropriately
with $\chi$ \cite{S-O}. And this makes a big difference, for the derivative of
$\chi$ cannot be isolated now in general (it requires an invertibility
condition on the Weyl spinor). Then, if we construct the corresponding
superenergy tensor (\ref{eq:T}), which still satisfies the dominant and other
required properties, it is impossible to check the divergence condition
{\it generically}. This is in accordance with the well-known problems of the
causal propagation of Rarita-Schwinger fields in curved backgrounds
(see e.g.\cite{AH}), which are reminiscent of the classical Velo-Zwanziger
causal paradox \cite{VZ} appearing when coupled to electromagnetic and
other external fields due again to the existence of some
invertibility conditions, see e.g.\cite{KT}.

\smallskip

To summarize, we have presented an extremely simple and fully
geometric argument for the causal
propagation of physical fields in arbitrary dimensions.
A sufficient condition, in many cases very easy to check explicitly,
for causal propagation is given in terms of the divergence of
an adequate superenergy tensor of the field. All our treatment is classical,
although it may be also useful in quantum aspects of the referred theories.
In this sense, notice that if any field may be described by tachyonic particles
then the divergence condition, which is related to the field equations, must be
violated. In general, we can choose to use any $T(A)$ with property
(\ref{eq:A0T0'}) that satisfies the divergence condition if this is helpful,
and in fact there usually exists an $s$-parameter family of such tensors for a
natural number $s$ \cite{S3}. Nevertheless, from (\ref{eq:w}) it is clear
that only the symmetric part of $T(A)$ matters and thus the divergence condition
refers to this intrinsically unique superenergy tensor (see also the discussion
in \cite{S3}). On the other hand, sometimes even the completely symmetric
tensors will not be unique (for instance in the case of (\ref{eq:T}), where
$\alpha$ and $\beta$ remain arbitrary), but in these cases the divergence
condition will hold for all of them and the conclusion about causal propagation
will of course be definitive. Furthermore, the condition (\ref{eq:div})
can be weakened, for it does not need to hold pointwise but only on integrated
form (even weaker conditions on the integral in (\ref{eq:wp2b}) can actually
be imposed for the conclusion (\ref{eq:w0}) to follow).
One can also formulate the conservation theorem for sets more
general than $\overline{D(S)}$, but we have restricted ourselves to
a physically interesting situation.

We believe that more sophisticated theorems for physical fields
can be found using superenergy tensors to define norms and
to do a priori estimates (as in \cite{CK}; with respect to this, see also
the excellent and very recent review \cite{KN}). The fact that it is always
possible to construct a superenergy tensor with the dominant property of any
field is fundamental here. Closer connections with the theory of symmetric
hyperbolic systems
need to be investigated because it may well happen that our divergence
condition (which is obviously connected to the field equations for $A$)
provide some useful criteria for the hyperbolic symmetrizability of some
field equations. All in all, we wish to stress the simplicity of our argument
and the universality of its application.

\smallskip

G.B. is grateful to The Swedish Natural Science Research Council (NFR)
for financial support.

\end{document}